\documentclass{cta-author}
\usepackage{lmodern}
\usepackage{amsmath,epsfig,multirow,amssymb,hyperref,makecell}
\begin{document}
\pagestyle{plain}
\title{Incorprating Prompt tuning for Commit classification with prior Knowledge}
\author{\au{Jiajun Tong$^{1\corr}$}, \au{Xiaobin Rui$^{1}$}}

\address{\add{1}{China University of Mining and Technology, School of Computer Science and Technology}
}

\begin{abstract}

  Commit Classification(CC) is an important task in software maintenance since it helps software developers classify code changes into different types according to their nature and purpose. This allows them to better understand how their development efforts are progressing, identify areas where they need improvement, and make informed decisions about when and how to release new versions of their software. 
  However, 
  existing methods are all discriminative models, usually with complex architectures that require additional output layers to produce class label probabilities, and they are task-specific, unable to learn features across different tasks.
  Moreover, they require a large amount of labeled data for fine-tuning, and it is difficult to learn effective classification boundaries in the case of limited labeled data. 
  To solve above problems, we propose a generative framework that Incorporating prompt-tuning for commit classification with prior knowledge (IPCK), which simplifies the model structure and learns features across different tasks. It can still reach the SOTA performance with only limited samples.
  Firstly, we proposed a generative framework based on T5. This encoder-decoder construction method unifies different CC task into a text2text problem, which simplifies the structure of the model by not requiring an extra output layer.
  Second, instead of fine-tuning, we design an prompt-tuning solution which can be adopted in few-shot scenarios with only limit samples. 
  Furthermore, we incorporate prior knowledge via an external knowledge graph to map the probabilities of words into the final labels in the speech machine step to improve performance in few-shot scenarios.
  Extensive experiments on two open available datasets show that our framework can solve the CC problem simply but effectively in few-shot and zeroshot scenarios, while improving the adaptability of the model without requiring a large amount of training samples for fine-tuning.
\end{abstract}

\maketitle
\thispagestyle{plain}
\section{Introduction}
\label{sec:intro}

Commit classification \citep{herivcko2023commit} is a software maintenance task that involves categorizing software source code commits into different classes or categories based on their content. In the context of software development, a commit is a set of changes made to a codebase, along with an associated message that describes the purpose and nature of those changes. 
It can affect the process of CI/CD and help developers efficiently manage and track changes in the codebase. Properly categorizing commits during software maintenance can bring many benefits. It enhances collaboration by providing context for changes made by developers. Clear categories simplify code reviews and help identify sources of bugs. This practice supports efficient knowledge transfer, helps track feature development, and enables targeted regression testing. Automated CI/CD processes benefit from triaging commits, triggering relevant tests. Committing classifications can also promote coding standards, facilitate documentation updates, and assist in planning refactorings. 
For example, Let's consider a scenario where a team is working on developing a web application. During a maintenance phase, a developer makes several commits to the codebase. 
\begin{table}[!ht]
  \centering
  \caption{Samples of Commit Classification}
  \resizebox{0.4\textwidth}{!}{\begin{tabular}{lll}
  \hline
~ & \textbf{Commit Message (Input Feature)} & \textbf{Label} \\ \hline

      \multirow{6}{*}{${Task}_{I}$} & \\ 
      ~ & Refactored database query for improved performance. & Adaptive \\ 
      ~ & Added new payment gateway feature. & Perfective \\ 
      ~ & Fixed critical bug in user authentication. & Corrective \\ 
      ~ & Optimized image rendering for faster loading. & Adaptive \\ 
      ~ & Updated user interface for a more intuitive experience. & Perfective \\ 
      ~ & ... & ... \\ \hline
      \multirow{6}{*}{${Task}_{II}$} & \\ 
      ~ & Added encryption to user passwords. & SECURE \\ 
      ~ & Updated external library to fix security vulnerability. & SECURE \\ 
      ~ & Removed exposed API endpoint. & SECURE \\ 
      ~ & Fixed XSS vulnerability in user profile page. & SECURE \\ 
      ~ & Refactored code for improved performance. & INSECURE \\ 
      ~ & Updated UI colors for a more vibrant look. & INSECURE \\ 
      ~ &  ... & ... \\ \hline
  \end{tabular}
  }
  \label{tab: Samples of Commit Classification}
\end{table}
As shown in Tab. \ref{tab: Samples of Commit Classification}, the commits are classified into different categories, such as Adaptive, Perfective, and Corrective. This allows the development team to quickly grasp the nature of each commit and its impact on the codebase, ultimately leading to a more efficient and productive development process.

Depending on the purpose, previous works can be divided into two branches, binary classification and multi-classification. 
Mokus et al. \citep{mockus2000identifying} first proposed the basis for multi-classification, through the statistical method.  This kind of approach is primarily rely on analyzing and modeling the statistical relationships and patterns in the data. And they often assume linear relationships or specific data distributions, which can limit their ability to capture complex patterns in the data. So that they might struggle to model non-linear relationships effectively.
Therefore, H{\"o}nel et al. \citep{honel2019importance} tried to use the machine learning method XGBoost to solve the CC problem, and introduced three additional features to improve the effect of CC. 
Similarly, to flag bug-related commits and bug reports, Zhou et al. \citep{zhou2017automated} proposed a binary classification solution which automated vulnerability identification system, through the ensemble learning algorithms.
Unlike many statistical methods that assume linear relationships, such nonlinear machine learning methods are effective at capturing complex relationships in the data, and can model intricate patterns without relying on explicit assumptions about data distribution.
In recent years, neural networks have achieved rapid development in various fields. They can automatically learn relevant features from the data, reducing the need for extensive manual feature engineering. With the help of neural network, Ghadhab et al. \citep{ghadhab2021augmenting} proposed a method based on neural network to build a multi-classification predictor, and released the dataset. 
While Lee et al. \citep{lee2021co} proposed a semi-supervised neural network based on pretrained CodeBERT to enhance the effect of prediction.
The ability of neural networks are especially useful when dealing with high-dimensional and unstructured data and learns to extract relevant features and make predictions without relying on intermediate manual processing steps.
\begin{figure*}[h]
  \centering
  \includegraphics[width=\linewidth]{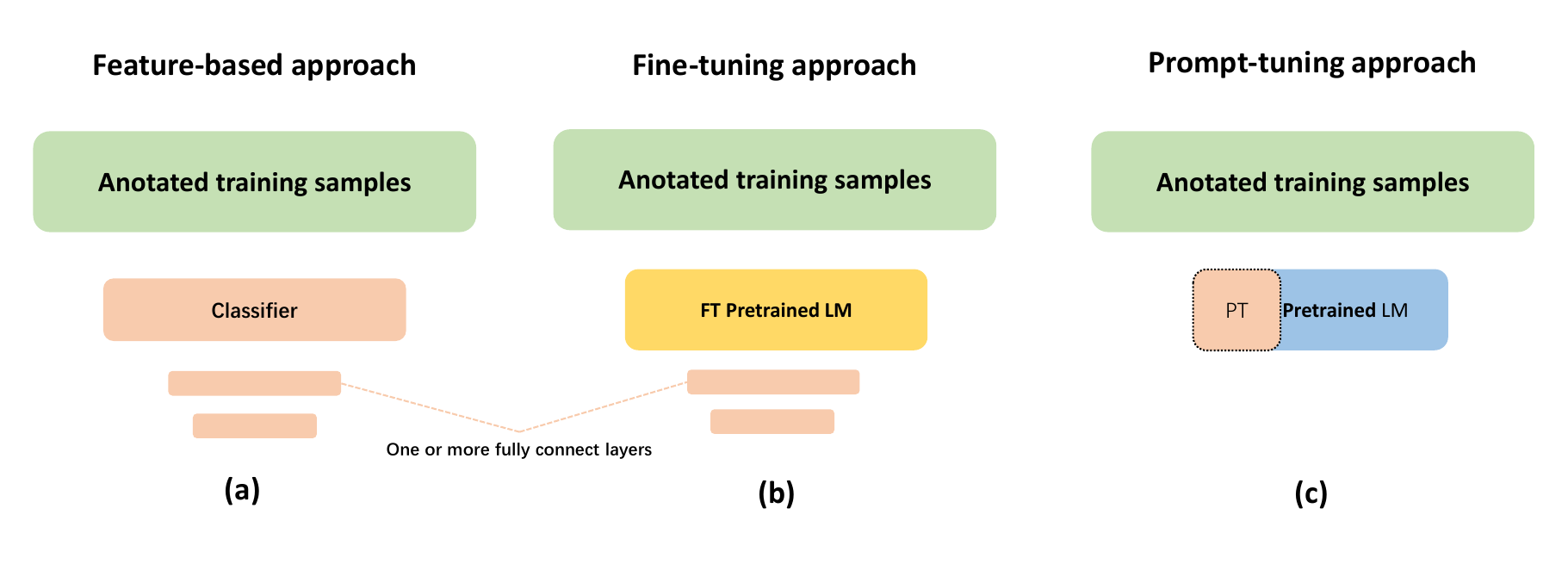}
  \caption{Comparison of different commit classification Paradigms. (a) The 1th and 2th Paradigms is Feature-based approaches which include (non-neural network and neural network) and require a lots of annotated samples for training. There both need an extra layer(e.g., a linear transformation) to obtain the outputs. (b) The 3rd paradigm is based on pretrained language model with a lots of training samples for fine-tuning, which also needs an extra layer(e.g., a linear transformation) to obtain the outputs. (c) The 4th Paradigms it prompt-based learning approaches that dose not need a lots of labeled samples for fine-tuning, and no need of extra layer for outputs.
  }
  \label{Comparison of different commit classification paradigms}
\end{figure*}
Overall, the development process of CC tasks are typically in line with the development process of NLP history. As shown in Fig.\ref{Comparison of different commit classification paradigms}(a)(b). The mainstream CC problems can be modeled in three paradigms. 
The first paradigm is feature-based approaches combined with linear models or neural networks. This type of method is less flexible and requires detailed extraction of features by experts. 
The second paradigm provides fine-tuning process on the pretrained model, which consists of unsupervised construction vectors from a large number of unlabeled data in the network and used as prior knowledge to enhance downstream tasks. 

However, existing methods often feature intricate architectures necessitating supplementary output layers for generating probabilities of class labels. Additionally, these models are confined to specific tasks, rendering them incapable of acquiring features across diverse tasks. 
For example, when dealing commit classification with vulnerability classification(multi-class) task, the softmax layer is often used as the output layer. A Softmax layer can convert the model's raw outputs (often called "logits") into a probability distribution for each class. This allows the output of the model to be directly interpreted as the probability that the input belongs to each class. while in security-related classification(binary-class) task, the sigmoid layer can be used as the output layer. The sigmoid layer converts the raw output of the model into a value between 0 and 1, which can be interpreted as the probability that the input belongs to a certain class. 

Moreover, they demand a substantial volume of labeled data for meticulous adjustment, and when confronted with limited labeled data, effectively learning classification boundaries becomes challenging.
The reason is that, fine-tuning approaches involve taking a pre-trained model (which has learned general language understanding from a large corpus of data) and adjusting its parameters to specialize it for a specific task. In the case of commit classification, like vulnerability classification, the model needs to learn task-specific patterns and nuances that are unique to that particular classification problem. These patterns might involve subtle relationships between words, phrases, and the programming codes. Since these patterns are specific to the task, a larger dataset is often needed to ensure that the model captures a wide range of these patterns and learns to generalize well to new examples. 



To bridge this gap, we present a novel generative framework in line with the fourth paradigm in Fig.\ref{Comparison of different commit classification paradigms}(c) known as "Incorporating Prompt-Tuning for Commit Classification with Prior Knowledge" (IPCK). This framework simplifies the model structure while enabling feature acquisition across a range of tasks. Impressively, it achieves state-of-the-art (SOTA) performance even with a restricted number of samples.
Firstly, we introduce a generative framework rooted in T5 architecture. This encoder-decoder approach unifies various Commit Classification tasks into a single text2text problem. Consequently, it streamlines the model's structure, eliminating the need for an additional output layer.
Secondly, different with from previous fine-tuning methodology, we devise a prompt-tuning solution ideal for few-shot scenarios. This involves adjusting a limited set of parameters with a modest sample pool.
Notably, we adopt external knowledge through a knowledge graph, which maps word probabilities to final labels during the speech machine phase. This incorporation of prior knowledge enhances the performance in few-shot scenario.
Extensive experiments on publicly available datasets, our framework demonstrates a capacity to adeptly address the Commit Classification issue, both in few-shot and zero-shot scenarios. Remarkably, it enhances the model's adaptability without requiring an extensive array of training samples for conventional fine-tuning.

The main contributions of this work can be summarized as follows:
\begin{itemize}
  \item [1)] For the first time, we formalize the commit classification task into a Text-To-Text  problem. Unlike previous discriminative methods, we proposed a generative framework based on T5 which can not only learn shared parameters between different tasks, but also simplifies the structure of the model by not requiring an extra output layer.
  \item [2)] Instead of fine-tuning, we design an prompt-tuning solution which can be adopted in few-shot scenarios, adjusting a small number of parameters, with only limit samples. Furthermore, we incorporate prior knowledge via an external knowledge graph to map the probabilities of words into the final labels in the speech machine step to improve performance in few-shot scenarios.
  \item [3)] We conduct extensive experiment to show that our framework can solve the CC problem simply but effectively in few-shot scenarios, and improve the adaptability of the model.
\end{itemize}

\section{Related work}

\noindent
\textbf{Language Models}
\noindent
Language models \citep{wang2022pre, yin2022survey, zhao2023survey} are a type of artificial intelligence model designed to understand, generate, and manipulate human language. The architecture of a language model determines how it processes and generates text. There are three main types of architectures commonly used in language models:  Decoder-Only, Encoder-Only, and Encoder-Decoder as shown in Tab.\ref{Tab: Language models under different architectures}.
The Decoder-Only architecture\citep{bautista2022scene, bevilacqua2022autoregressive}, also known as a generative model, is designed to generate text given some context or prompt. It takes a partial sequence as input and generates the subsequent tokens one by one. This type of architecture is commonly used for tasks like language generation, text completion, and machine translation. An example of the Decoder-Only architecture is the GPT-2 model, which is specifically designed for text generation.
The Encoder-Only\citep{salazar2019masked, song2019mass} architecture, also known as an autoregressive model, focuses on understanding and encoding input text. It takes a sequence of tokens as input and processes them through multiple layers, capturing contextual information at each step. This architecture is often used for tasks like language understanding, sentiment analysis, and text classification. The most well-known example of an Encoder-Only architecture is the GPT (Generative Pre-trained Transformer) series of models developed by OpenAI.
The Encoder-Decoder\citep{sanh2021multitask,raffel2020exploring} architecture combines the concepts of both the Encoder and the Decoder. It is commonly used for sequence-to-sequence tasks such as machine translation, text summarization, and chatbot interactions. The Encoder processes the input sequence and produces a fixed-size representation, often referred to as the context vector. This context vector is then fed into the Decoder, which generates the output sequence token by token. The Transformer architecture, on which model like T5 can be used in an Encoder-Decoder configuration for various tasks.
Since the models under decoder-only architecture are often very large, not easy to migrate and tune. Encoder-only models like BERT, usually requires an additional classification layer, which is inconsistent with our original intention of simplifying the model architecture. 
So we here adopt the Encoder-decoder model T5 as the backbone of our proposed framework.

\begin{table}[ht]
  \centering
  \resizebox{0.5\textwidth}{!}{\begin{tabular}{lll}
  \hline
      \textbf{Language Model Type} & \textbf{Architecture}  & \textbf{Examples} \\ 
      Autoregressive LM & Decoder-Only & GPT, GPT-2 \\
      Masked LM & Encoder-Only & BERT, RoBERTa \\ 
      \textbf{Text-to-Text} & \textbf{Encoder-Decoder} &\textbf{T5} \\ \hline
  \end{tabular}
  }
  \caption{Language models under different architectures}
  \label{Tab: Language models under different architectures}
\end{table}

\noindent
\textbf{Prompt-tuning}
Prompt-tuning(PT) \citep{lester2021power,liu2022p,gu2021ppt} can make full use of the advantages of language models, and is currently a popular research direction on NLP. 
%
Nowadays, more and more scholars have begun to explore to apply the PT method to the task of classification.
\citep{chen2022knowprompt} propose a knowledge-aware prompt adjustment method (KnowPrompt) with co-optimization that can incorporate knowledge between relation labels into prompt adjustment for relation extraction.
\citep{han2022ptr}  We propose a method that encodes prior knowledge in classification tasks into rules, then designs sub-cues according to the rules, and finally combines the sub-cues to handle the classification task.
However, there is no work related to prompt-tuning for commit classification.
To bridge this gap, we present a novel generative framework Incorporating prompt-tuning for commit classification with prior knowledge (IPCK), which simplifies the model structure while enabling feature acquisition across a range of tasks. Impressively, it achieves state-of-the-art (SOTA) performance even with a restricted number of samples.

\noindent
\textbf{Commit Classification} 
The commit classification task \citep{zhou2021finding,wu2022enhancing} has gained significance due to its potential to enhance the efficiency and quality of software development processes. By automatically categorizing commits, developers and project managers can gain insights into the nature of code changes, track progress, identify bug fixes, and even automate certain aspects of the development workflow. This task is particularly important in large software projects where numerous code changes occur regularly.
Existing commit classification approaches can be divided into three types according to the NLP paradigm: 
1. Feature-based approaches\citep{sabetta2018practical,mariano2019feature,levin2017boosting} which include (non-neural network and neural network) and require a lots of annotated samples for training. There both need an extra layer(e.g., a linear transformation) to obtain the outputs. 
2. the 3rd paradigm is based on pretrained language model \citep{lee2021co, ghadhab2021augmenting} with a lots of training samples for fine-tuning, which also needs an extra layer(e.g., a linear transformation) to obtain the outputs. 
However, to the best of our knowledge, there is no existing approaches in line with the fourth paradigm for commit classification tasks.

\section{METHODOLOGY}

\noindent
\textbf{Problem definitions}
The problem we address in this study it to predict the label $y$ of a given commit message $x$. Specifically, Given a few labeled commits $(x_i, y_i)$, we adopt the commonly used $N$-way $K$-shot strategy to train the model, where $N$ represents the number of commit classes and $K$ represents the quantity of annotated commits for each class. Each task consists of a support set $S$ that contains $N \times K$ support instances and a query set $Q$. We train a classifier using the support set $S$ and evaluate its performance on the query set $Q$. In this setting, a higher value of $N$ and a lower value of $K$ indicate a more challenging task, as the classifier needs to generalize well with limited annotated data.

\noindent
\textbf{Overview}
In this section, we present our proposed method Incorporating Prompt tuning for Commit classification with prior Knowledge(IPCK). We first give an overall view for prompt tuning and elaborate on the construction of knowledge-enhanced verbalizer. Furthermore, we introduce how to incorporate contrastive learning as a preprocessing method to help the model focus on high-level features. Finally we outline the training and inference process of our proposed method.

Let $M$ be a pretrained language model. Our goal is to assign the class label $y$ from a set of labels $Y$ to an input sequence $x = (x_0, x_1, ..., x_n)$. To fully explore the knowledge in the pretrained model, we transform the commit classification task into a Masked Language Modeling(MLM) problem through prompt-tuning.
\begin{figure}[h]
  \centering
  \includegraphics[width=\linewidth]{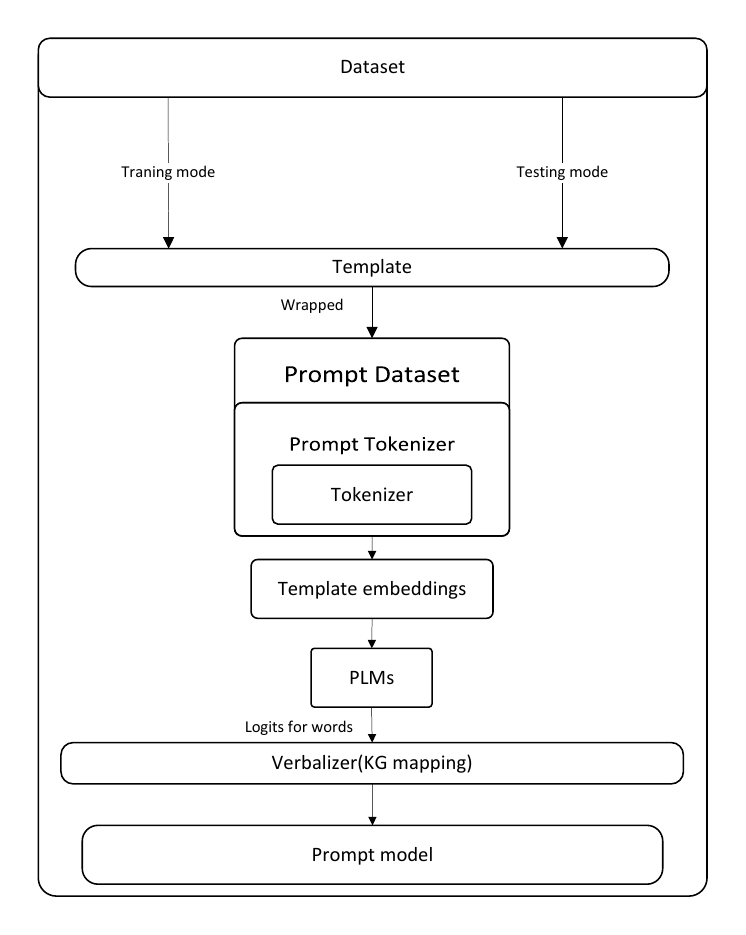}
  \caption{Overview of IPCK: IPCK can be divided into two stages. The first stage is training mode, }
  \label{Overview of IPCK}
\end{figure}
Notably, the prompts used in prompt-tuning feature an empty slot to be filled with $\mathsf{MASK}$, either in the middle or at the end of the prompt. These template words can be virtual words represented by numeric IDs or continuous vector. Some prompting methods even directly generate continuous vectors. The number of [$X$] slots and [$\mathsf{MASK}$] slots can be adjusted according to specific mission requirements.

In our case, the input sequence is wrapped with a template, which is a piece of natural language text. For instance, if we aim to classify following commit message $x =$ ``Clean up the man page, remove the reference to scrub mode" into the label "SECURE" or "INSECURE", we wrap it as follows:
\begin{equation}
  x_{\text{prompt}} = x~\text{This commit is}~[\mathsf{MASK}].
  \end{equation}
Next, the language model $M$ calculates the probability of each word $v$, which fills the [$\mathsf{MASK}$] token as $P_M([\mathsf{MASK}] = v|x_{prompt})$. To map the word probabilities to label probabilities, we define the Verbalizer as a mapping function, which maps words in the vocabulary $V$ into the label space $Y$. Then we use the subset of $V$ that maps to a particular label $y$, which can be denoted as $V_y$. Therefore,  $P(y|x_{\text{prompt}})$ which represents the probability of commit class $y$ can be calculated as follows:
\begin{equation}
  P(y|x_{\text{prompt}}) = g(P_M([\mathsf{MASK}] = v|x_{\text{prompt}})|v\in V_y),
  \end{equation}
  where $g$ represents a function that transforms the probability of label words into the probability of the commit class. 

\noindent
\textbf{Commit Tokenizer}
For commit tokenization, we leverage T5 to transform a commit message (input) into a sequence of tokens (output) that accurately represents the constituent elements of the commit.  T5 (Text-To-Text Transfer Transformer) is a state-of-the-art language model developed by Google Research that excels in a wide range of natural language processing tasks by reframing them into a unified text-to-text framework. In the context of commit tokenization, T5 can be leveraged to efficiently process and tokenize commits in version control systems like Git. T5's architecture is based on the Transformer model, which utilizes self-attention mechanisms to capture contextual relationships within a text sequence. This allows T5 to understand dependencies between tokens and generate high-quality tokenizations for various types of text, including commit messages.

By leveraging T5's text-to-text capabilities, we can create a robust and adaptable commit tokenizer that handles a wide variety of commit message formats and structures. 

\noindent
\textbf{Knowledgable Verbalizer}

To construct a knowledgable verbalizer, we consider it as an N-way K-shot scenario, where we have $N$ categories denoted as ${C_1, C_2, ..., C_N}$. Instead of directly using class names as label words, we aim to incorporate more semantic information into the verbalizer. Inspired by the concept of knowledgable Prompt \citep{deng2022simemotion}, we utilize an external knowledge graph called ``Related Words" \footnote{\url{https://relatedwords.org}} to expand the set of rich label words for each class $C_i$ based on their class name.

We leverage the Related Words knowledge graph to compute the similarity between the predicted [$\mathsf{MASK}$] tokens and the nodes in the knowledge graph. We formalize their the similarity by using the Jaccard similarity coefficient as follow:

\begin{equation}
J(y_e,V_y) = \frac{|y_e \cap Y_e|}{|y_e \cup Y_e|}
\end{equation}

Here, $y_e$ represents the embeddings of the [$\mathsf{MASK}$] tokens, and $Y_e$ represents the embeddings of candidate nodes in the knowledge graph. By computing the Jaccard similarity coefficients, we obtain the top $N^{KG} = 20$ relevant words associated with each class. These candidate words include synonyms and words highly related to the class name. For instance, for the class ``SECURE," the candidate words may include ``safe", ``ensure", ``fix", ``assurance", and so on.

We merge the candidate words into a single prototype for each class by averaging the embeddings of the candidate words. This process yields $N$ synthesized continuous label word embeddings, denoted as $W^{(0)} \in \mathbb{R}^{N \times D}$. The matrix $W^{(0)}$ represents the rough semantic meaning of the $N$ classes and serves as the initial set of label word embeddings. These embeddings will be further fine-tuned in the subsequent module.

\noindent
\textbf{Training and inference}
For training stage, given a input $x$, the target is to obtain the probability distribution over labels $P(y|x_p)$. We calculate the dot product between the $x$ embedding $h$ and task-adapted label word embeddings $w_y$ to predict the probability score by using the softmax function:
\begin{equation}
  P(y|x_p) = \frac{\exp(w_y \cdot h)}{\sum_{w \in W}\exp(w \cdot h)}
\end{equation}
In addition to the classification loss, we also incorporate language modeling as an auxiliary task to address the limited training samples. We adopt the standard token-level cross-entropy loss as the loss function. The loss will be calculated based on the difference between the predicted output label sequence and the actual label sequence:
\begin{equation}
  CE(y,\hat{y}) = -\sum_{i = 1}^{N_c}y_i \cdot log(\hat{y_i})
\end{equation}
where $y$ is the true or ground truth distribution of the target labels. and $\hat{y}$ represents the predicted or estimated distribution of the target labels. $N_c$ is the number of classes or categories in our classification task. While
$y_i$ represents the $i$-th element of the true distribution $y$, and $\hat{y_i}$ represents the $i$-th element of the predicted distribution $\hat{y}$.

During inference, given a new input commit message, we apply the prompt-verbalizer pair $(p, v)$ to generate the prompt template $x_{prompt}$ and compute the probability distribution over labels $P(y|x_{prompt})$ using the trained language model $M$. The class with the highest probability is assigned to the commit.
\section{Experiment}
In this section, we first introduce the dataset and the experiment setup. Then we compared with different baselines to show the superiority of our model to the current state-of-the-art(SOTA) models.
We further evaluate the performance of our model with different numbers of samples as the training set to illustrate the advantages of the generative based framework in the case of fewshot scenario.
Finally, we visualized the performance of the model on the two data sets to show the comprehensive performance of the model and the detailed performance of the model on each target.
Our experimental environment is as follows: We limit the early stop function up to 10 epochs after no improvements on Acc. All models are trained on Gpushare Cloud a leading GPU Cloud service provider from China with GPI 24GB Nvidia 3090 and Intel(R) Xeon(R) CPU E5-2683 v4 with 40GB memory. 


  \subsection{Experimental Setup} 

  \noindent
  \textbf{Dataset}
  \noindent
  We evaluate two publicly available datasets to demonstrate the effectiveness of our proposed model.  Since ``comment'' is the common attribute on the two datasets, we want to demonstrate our model as a simple and general approach. Therefore, we choose comments as our only input. Ghadhab et al. \citep{ghadhab2021augmenting} combined three datasets \citep{mauczka2015dataset, alomar2019can} which collect commits from open-source projects that cover several domains (e.g., databases, programming languages and integration frameworks), with 1,793 annotated commits and three categories for software maintenance activities identification. We denoted it as Dataset I\footnote{\url{https://zenodo.org/record/4266643\#.X6vERuLPxPY}} and present the data characteristics in Tab.\ref{tab: Data charcteristics of Dataset I} and Tab.\ref{tab: Samples of Dataset I}. 
  Lee et al. \citep{lee2021co} create a dataset from RA-Data \citep{reis2021ground}, which consist of 3,765 positive samples and roughly 6,300 negative samples from 910 repositories with two classes SECURE and INSECURE. We denoted it as Dataset II\footnote{\url{https://github. com/davidleejy/wnut21-cotrain}} and present the data characteristics in Tab.\ref{tab: Data characteristics of Dataset II} and Tab.\ref{tab: Samples of Dataset II}.

  \begin{table}[!ht]
    \centering
    \caption{Data charcteristics of Dataset I(2 labels)}
    \begin{tabular}{ll}
      Data Charcteristics of Dataset I & ~ \\ 
      \hline
      Category label & Number of Instances \\ 
      \hline
      Positive & 6347 \\ 
      Negative  & 6347 \\ 
      \hline
      Total & 12694 \\ 
      \hline
    \end{tabular}
    \label{tab: Data charcteristics of Dataset I}
\end{table}

  \begin{table}[h]
    \centering
    \caption{Data Charcteristics of Dataset II(3 labels)}
    \begin{tabular}{ll}
        Data Charcteristics of Dataset II & ~ \\ \hline
        Category label & Number of Instances \\ \hline
        Corrective & 600 \\
        Adaptive & 590 \\ 
        Perfective & 603 \\ 
        \hline
        Total & 1793 \\ 
        \hline
    \end{tabular}
    \label{tab: Data characteristics of Dataset II}
  \end{table}
  We split the datasets with 70\% train, 15\%test, and 15\% val. And as our method is focused on few-shot and zeroshot scenarios, we follow the general N-way K-shot sampling strategy to extract training samples from the training dataset.

  \noindent
  \textbf{Baselines}
  Although there are many articles on CC, only a small amount of work is based on public datasets. In order to fairly compare the performance of our model, we choose compare approaches with experimental methods based on public datasets. 
  Further, in order to reflect the effectiveness of different component in our model, we have prepared four ablation versions of IPCK as our baselines. These methods are as follows:
  \begin{itemize}
    \item CC-ds: It is a binary classification task. Lee et al. \citep{lee2021co} treated code changes and commit messages as two different views and used CodeBERT and RoBERTa to process them, respectively. They applied co-training to train the two models jointly.
    \item DNN@BERT+Fix\_cc: It is a multiclass classification task. Ghadhab et al. \citep{ghadhab2021augmenting} introduced a DNN model which concatenates the BERT-based word embeddings of commit messages and source code changes.
    \item  $\mathsf{{IPCK}^{M}_{T5}}$  In order to compare the impact of different pre-training models on IPCK performance, we propose a T5-based version, and utilize the Manual(M) verbizer that does not introduce prior knowledge.
    \item  $\mathsf{{IPCK}^{K}_{T5}}$ This is an ablation version of IPCK, we leverage the Knowledgable(K) verbilzer which incorporate  prior knowledges from external knowledge graph, but trained on T5.
    \item Similar to $\mathsf{{IPCK}^{M}_{T5}}$, we replace the pre-trained model with flan-T5, which is fine-tuned on more than 1000 additional tasks, covering more language, based on the same number of parameters.
    \item $\mathsf{{IPCK}^{K}_{flan-T5}}$ Similar to  $\mathsf{{IPCK}^{K}_{T5}}$ we leverage the Knowledgable(K) verbilzer which incorporate  prior knowledges from external knowledge graph, but replace the pre-trained model with flan-T5.
   
  \end{itemize}

  \noindent
  \textbf{Implementation and Hyperparameters}
  We provide the hyper-parameter values for IPCK as follows: We utilize the default parameters from Huggingface \footnote{\url{https://huggingface.co}} to ensure fairness. Specifically, we introduce AdamW as the optimizer function, with a learning rate of 1e-5 and the batch size is set to 64. And we limit the early stop function up to 10 epochs after no improvements on Acc. All models are trained on 2 Nvidia Tesla A100 with 16GB memory.  detailed as follow table
  \noindent
  \textbf{Metrics} 
  Evaluating the performance of commit classification models is essential to assess their effectiveness in accurately categorizing code changes. In this section, we present the evaluation metrics employed to measure the performance of our commit classification model. These metrics provide valuable insights into the model's ability to distinguish between different types of commits and its overall effectiveness. 
  \begin{itemize}
    \item \textbf{Accuracy} is a fundamental metric that gauges the proportion of correctly classified commits among all commits in the dataset. It provides an overall assessment of the model's performance and helps determine its general classification prowess.
    \item  \textbf{Precision} measures the proportion of true positive commits among all commits classified as a specific class. In the context of commit classification, precision denotes the model's ability to correctly identify commits belonging to a particular category, such as bug fixes, feature additions, or documentation updates. A high precision score indicates a low number of false positives.
    \item \textbf{Recall}, also known as sensitivity or true positive rate, evaluates the ability of the model to correctly capture all positive instances of a specific class. It calculates the proportion of true positive commits among all commits that truly belong to that class. A high recall score implies that the model can effectively identify a significant portion of the commits belonging to the target class.
    \item The \textbf{F1-Score} is a harmonic mean of precision and recall, providing a balanced measure of the model's performance. It takes both false positives and false negatives into account, making it particularly useful when dealing with imbalanced datasets. A high F1-Score indicates a well-balanced trade-off between precision and recall.
  \end{itemize}
  By employing these evaluation metrics, we can rigorously analyze and interpret the results of our commit classification model and make informed decisions about its practical deployment in real-world software development scenarios.
\subsection{Performance Comparison}
\label{sec: Performance Comparison}
In order to illustrate the SOTA performance of our proposed method and verify the adaptability on two datasets. We compared with the current state-of-the-art methods on two open source datasets.
From Tab \ref{tab: Performance comparison on Dataset I}, we can see that our method $\mathsf{{IPCK}^{K}_{T5}}$ based on the knowledgable verblizer and T5 model has achieved the best results. It is worth noting that our method achieves more than 90\% in all three indicators when only using the message feature, and does not require an additional output layer.
It is understandable that the method $\mathsf{{IPCK}^{K}_{flan-T5}}$ incorporating prior knowledges and bigger weigthed pretrained model did not achieve the best results. This may be because FLAN-T5 is mainly based on large Large-scale natural language corpus training, and CC problems include not only natural language, but also a large number of professional terms and programming languages. Therefore, in the pursuit of higher accuracy, these features need to be further considered.
On the dataset II, the method $\mathsf{{IPCK}^{K}_{T5}}$ based on the knowledgable verblizer achieved the best results from \ref{tab: Performance comparison on Dataset II}.
Overall, from the two experiments we can see that our method achieves SOTA performance on both datasets. 
This shows that our method can simply and effectively improve the performance on CC tasks. And has good adaptability.
\begin{table}[h]
  \centering
  \caption{Performance comparison on Dataset I(2 labels)} 
  \begin{tabular}{lccc}
      \textbf{Method} & \textbf{Precision} & \textbf{Recall} & \textbf{F1-score} \\ \hline
      Training samples 7000+ \\ 
      CR-ds & 84.0 & 84.1 & 83.9 \\
      $\mathsf{{IPCK}^{M}_{T5}}$ & 89.65 & 89.45 & 89.51 \\
      $\mathsf{{IPCK}^{M}_{flan-T5}}$ & 89.41 & 88.89 & 89.00 \\
      $\mathsf{{IPCK}^{K}_{T5}}$ & \textbf{90.15} & \textbf{90.15} & \textbf{90.15} \\ 
      $\mathsf{{IPCK}^{K}_{flan-T5}}$ & 88.68 & 88.33 & 88.42 \\ \hline

  \end{tabular}
  \label{tab: Performance comparison on Dataset I}
\end{table}
\begin{table}[h]
  \centering
  \caption{Performance comparison on Dataset II(3 labels)} 
  \begin{tabular}{lccc}
      \textbf{Method} & \textbf{Precision} & \textbf{Recall} & \textbf{F1-score} \\ \hline
      Training samples 1200+ \\ 
      DNN@BERT+Fix\_cc & 80.0 & 79.7 & 79.7 \\  
      $\mathsf{{IPCK}^{M}_{T5}}$ & 80.95 & 80.48 & 80.50 \\
      $\mathsf{{IPCK}^{M}_{flan-T5}}$ & 81.01 & 80.48 & 80.46 \\
      $\mathsf{{IPCK}^{K}_{T5}}$ & \textbf{83.48} & \textbf{83.46} & \textbf{83.43} \\ 
      $\mathsf{{IPCK}^{K}_{flan-T5}}$ & 82.60 & 82.53 & 82.54 \\ \hline

  \end{tabular}
  \label{tab: Performance comparison on Dataset II}
\end{table}
\subsection{The Evaluation on Fewshot scenario}
In order to verify the feasibility of our proposed method in the fewshot scenario, we compared our ipck ablation version in the case of \{5, 10, 15, 20, and 50\} shots, to observe the performance of the model.

\begin{figure*}[ht]
  \centering
  \includegraphics[width=\linewidth]{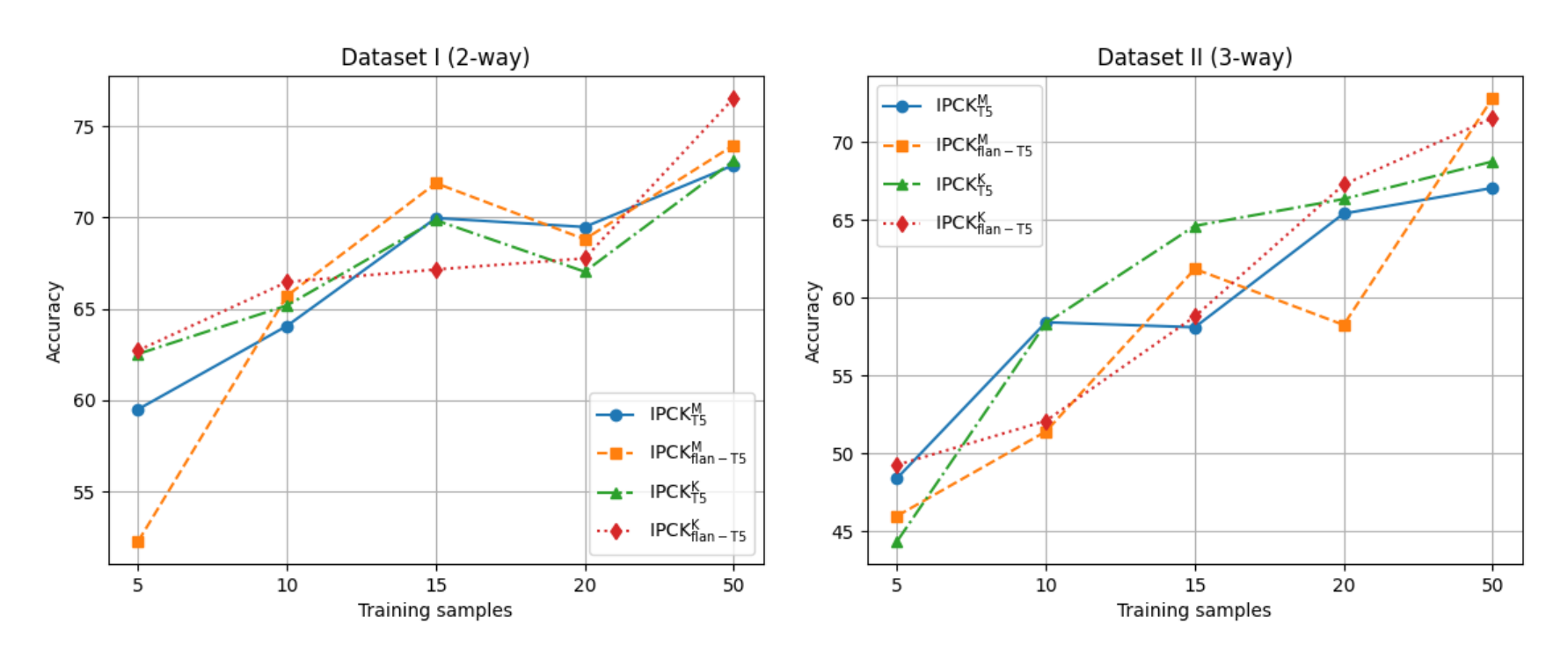}
  \caption{Fewshot performance on different datasets
  }
  \label{fig: Fewshot perfermance}
\end{figure*}

From Tab. \ref{tab: Performance on Fewshot scenerios}, we can see that the performance of IPCK has been significantly improved with the increase of data volume. At 50 shots, it achieves more than 70\% performance on both datasets.
We have highlighted the best performance. As expected, the IPCK based on the flan-T5 model trained with more training parameters achieved better results, which shows that the performance of the model is proportional to the training parameters.
Further, by observing the performance of $\mathsf{{IPCK}^{M}_{flan-T5}}$ and $\mathsf{{IPCK}^{K}_{flan-T5}}$, we found that the application The $\mathsf{{IPCK}^{K}_{flan-T5}}$ with the knowledgable verblizer achieved the best results on dataset I, and slightly lower than $\mathsf{{IPCK}^{M}_{flan-T5}}$. This may be because the training data set is randomly divided, and there is information that is not easily identifiable.
In order to observe the impact of different modules on the overall performance of the fewshot scene, we visualized the ACC in different shot situations on two datasets.
From Fig. \ref{fig: Fewshot perfermance} we can see that $\mathsf{{IPCK}^{K}_{flan-T5}}$ has achieved the most stable effect, while other models are different degree of fluctuation.

Generally speaking, we can conclude that our model can be effectively applied to the fewshot scene, and using the T5 model with more parameters and introducing external knowledgable is helpful to improve the performance of the model in the fewshot scene.
\begin{table*}[ht]
  \begin{center}
  \caption{IPCK's performance on Fewshot scenerios} 
 
  \begin{tabular}{clcccccccccccc}
    \hline
    \multirow{2}{*}{Shots} & \multirow{2}{*}{Methods} & \multicolumn{3}{c}{Dataset I} & \multicolumn{3}{c}{Dataset II} \\
    ~ & ~ & P & R & F & P & R & F \\ \hline
    \multirow{4}{*}{5} & $\mathsf{{IPCK}^{M}_{T5}}$ & 57.14 & 59.49 & 57.61 & 49.88 & 48.42 & 45.44 \\
   ~& $\mathsf{{IPCK}^{M}_{flan-T5}}$ & 57.55 & 52.25 & 52.79 & 46.33 & 46.00 & 41.35 \\
   ~& $\mathsf{{IPCK}^{K}_{T5}}$ & 52.90 & 62.52 & 48.87 & 48.22 & 44.37 & 40.29 \\ 
   ~& $\mathsf{{IPCK}^{K}_{flan-T5}}$ & 57.79 & 62.71 & 50.56 & 49.68 & 49.27 & 47.60 \\ \hline
    \multirow{4}{*}{10} & $\mathsf{{IPCK}^{M}_{T5}}$ & 61.51 & 64.06 & 58.79 & 58.39 & 58.42 & 57.88 \\
    ~& $\mathsf{{IPCK}^{M}_{flan-T5}}$ & 64.09 & 65.71 & 63.82 & 55.78 & 51.42 & 48.79 \\
    ~& $\mathsf{{IPCK}^{K}_{T5}}$ & 63.37 & 65.18 & 60.17 & 59.05 & 58.36 & 57.62 \\ 
    ~& $\mathsf{{IPCK}^{K}_{flan-T5}}$ & 65.78 & 66.47 & 61.20 & 53.93 & 52.10 & 49.58 \\ \hline
    \multirow{4}{*}{15} & $\mathsf{{IPCK}^{M}_{T5}}$ & 69.09 & 69.97 & 69.02 & 61.27 & 58.10 & 57.61 \\
    ~& $\mathsf{{IPCK}^{M}_{flan-T5}}$ & 71.26 & 71.88 & 70.44 & 62.37 & 61.84 & 61.26 \\
    ~& $\mathsf{{IPCK}^{K}_{T5}}$ & 68.95 & 69.85 & 68.86 & 64.91 & 64.60 & 64.06 \\ 
    ~& $\mathsf{{IPCK}^{K}_{flan-T5}}$ & 66.23 & 67.15 & 66.42 & 60.45 & 58.79 & 57.57 \\ \hline
    \multirow{4}{*}{20} & $\mathsf{{IPCK}^{M}_{T5}}$ & 68.98 & 69.48 & 69.14 & 65.21 & 65.39 & 65.08\\
    ~& $\mathsf{{IPCK}^{M}_{flan-T5}}$ & 67.72 & 68.81 & 67.26 & 59.96 & 58.25 & 57.80 \\
    ~& $\mathsf{{IPCK}^{K}_{T5}}$ & 66.81 & 67.03 & 66.91 & 66.46 & 66.32 & 66.35 \\ 
    ~& $\mathsf{{IPCK}^{K}_{flan-T5}}$ & 67.25 & 67.76 & 63.61 & 67.96 & 67.25 & 66.71 \\ \hline
    \multirow{4}{*}{50} & $\mathsf{{IPCK}^{M}_{T5}}$ & 73.69 & 72.87 & 73.14 & 69.37 & 67.03 & 66.44 \\
    ~& $\mathsf{{IPCK}^{M}_{flan-T5}}$ & 73.65 & 73.96 & 73.76 & \textbf{73.66} & \textbf{72.79} & \textbf{72.59} \\
    ~& $\mathsf{{IPCK}^{K}_{T5}}$ & 72.68 & 73.11 & 71.72 & 71.00 & 68.73 & 68.64 \\ 
    ~& $\mathsf{{IPCK}^{K}_{flan-T5}}$ & \textbf{76.26} & \textbf{76.54} & \textbf{76.34} & 72.60 & 71.52 & 71.71 \\ \hline

  \end{tabular}
  \label{tab: Performance on Fewshot scenerios}
\end{center}
\end{table*}
\subsection{Result Visualization}

\begin{figure*}[ht]
  \centering
  \includegraphics[width=\linewidth]{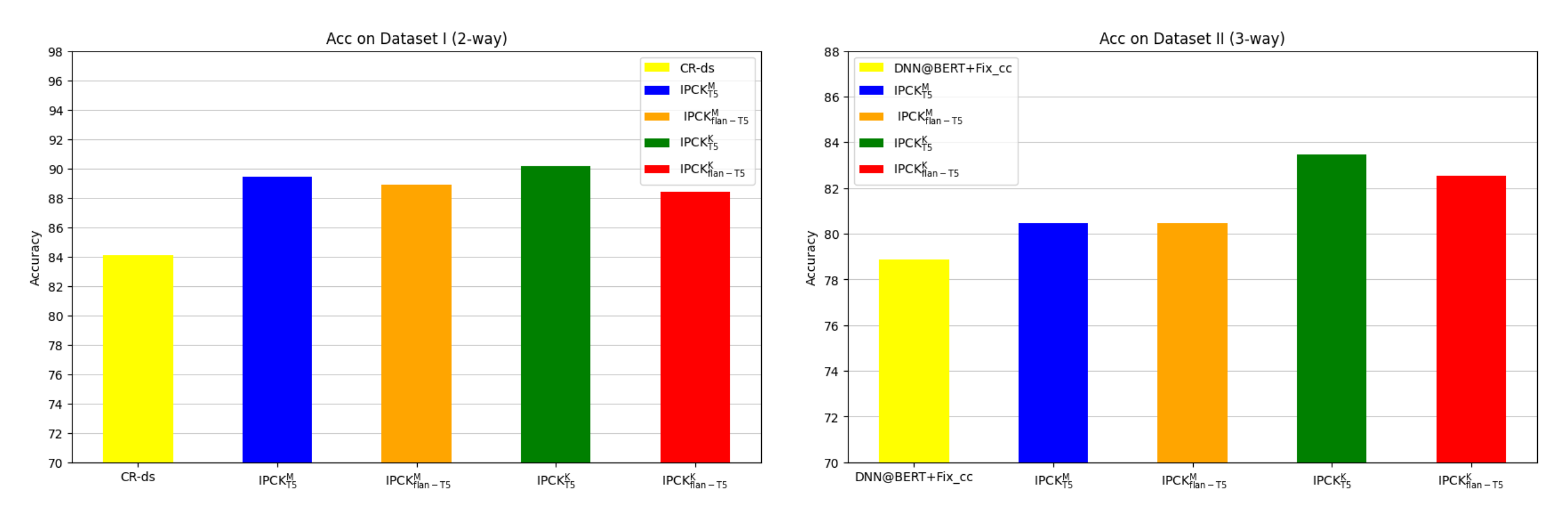}
  \caption{Acc performance on different datasets.
  }
  \label{Acc performance on different datasets}
\end{figure*}

\begin{figure*}[ht]
  \centering
  \includegraphics[width=\linewidth]{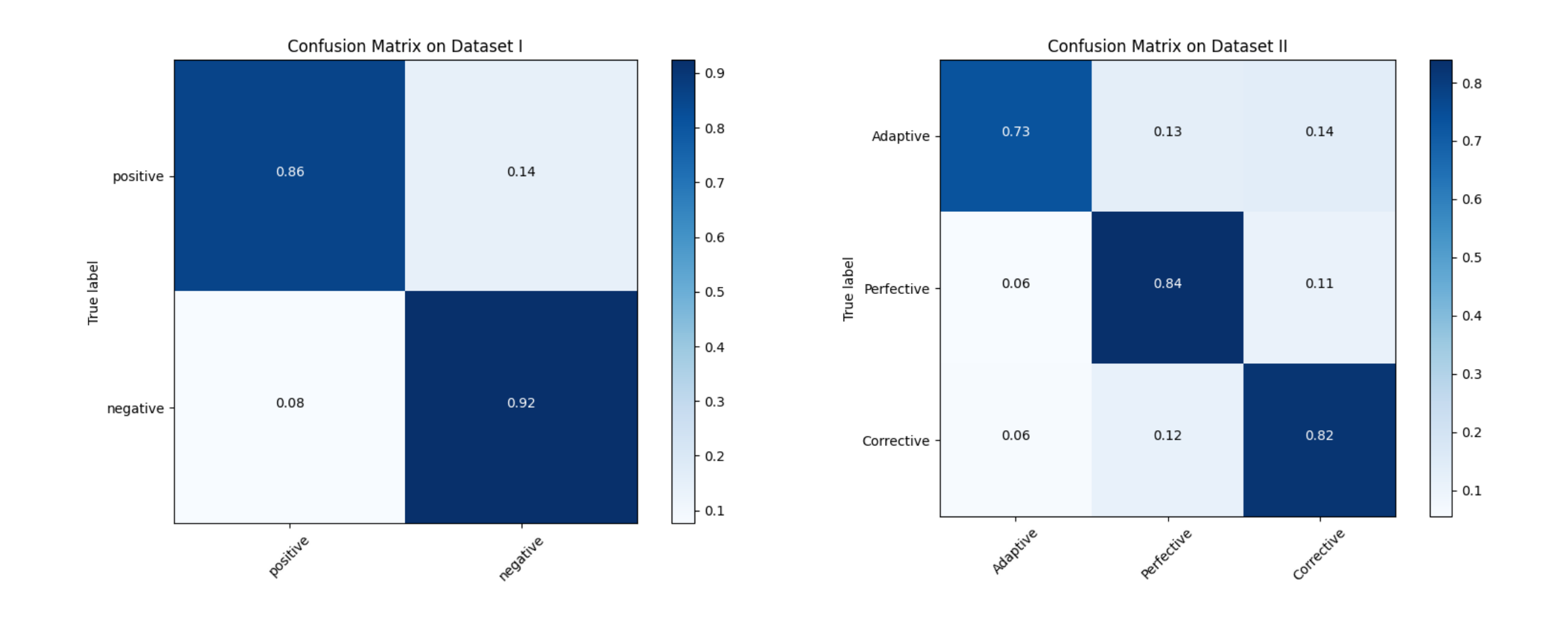}
  \caption{Confusion matrix on different datasets. 
  }
  \label{Confusion matrix}
\end{figure*}

In order to show the details of our model's prediction effect more specifically, we visualized the results of IPCK on two data sets.
First of all, we show the ACC between our proposed method and the previous method on two data sets, as shown in Fig.\ref{Acc performance on different datasets}. From the figure, we can see that our method has achieved the best results on both data sets Very high ACC, which shows that, overall, our proposed method can effectively improve the performance of CC problem SOTA.
First of all, we show the ACC between our proposed method and the previous method on two data sets, as shown in Figure xxx. From the figure, we can see that our method has achieved the best results on both data sets Very high ACC, which shows that, overall, our proposed method can effectively improve the performance of CC problem SOTA.
we can see from Fig.\ref{Confusion matrix} that on Dataset II, our model achieves 73\%, 84\% and 82\% accuracy in the classification of adaptive, perfective and corrective targets. This is understandable because the three-category target may require more features To learn, and we only use message as a single input feature here.

\section{Conclusion} 
Commit Classification(CC) is an important task in software maintenance. Existing models need lots of manually labeled data for fine-tuning process, and not easy to adapt different kinds of datasets. Therefore, it is still challenging to employ generative models on commit classification tasks in few-shot and zeroshot scenarios. In this work, we propose a framework named Incorporating Prompt tuning for Commit classification with prior Knowledge(IPCK), which does not require lots of annotated samples and is easy to adapt to different tasks. 
The experimental results on two available datasets prove that IPCK is a simple but effective method to solve CC task on few-shot or even zeroshot scenarios, and effectively improve the adaptability of the model without requiring a large amount of data for fine-tuning.
In future work, we are going to explore soft prompts to automatically summarize effective prompt tokens by compressing information from a large number of existing datasets. Further, simplify the tuning process under different commit classification standards.

\section*{Data availability}
The raw/processed data required to reproduce these findings cannot be shared at this time as the data also forms part of an ongoing study.

\section*{Conflicts of interest}
The authors declare that there is no conflict of interest regarding the publication of this article.

\section*{Funding statement}
This work was supported by the National Natural Science Foundation of China (No.61876186).

\bibliographystyle{iet}
\bibliography{iet}
\end{document}